\documentclass[pra,twocolumn,showpacs,preprintnumbers,amsmath,amssymb,longbibliography,
               superscriptaddress]{revtex4-2}

\usepackage{amsmath,amssymb,amsfonts}
\usepackage{graphicx}
\usepackage{hyperref}
\usepackage{booktabs}
\usepackage{bm}
\usepackage{tikz}
\usepackage{pgfplots}
\pgfplotsset{compat=1.18}

\begin{document}

\title{{Singlet-Like Correlations: Equal Peak Work, Unequal Robustness}}

\author{Karl Svozil\,}
\email{karl.svozil@tuwien.ac.at}
\homepage{http://tph.tuwien.ac.at/~svozil}
\affiliation{Institute for Theoretical Physics, TU Wien,
  Wiedner Hauptstra{\ss}e 8-10/136, 1040 Vienna, Austria}

\date{\today}

\begin{abstract}
{Initial system--environment correlations are a thermodynamic resource, enabling work extraction through their erasure. We compare three representative \emph{singlet-like, rotationally covariant} correlation laws---a local classical benchmark, the quantum cosine law, and an idealized stronger-than-quantum step law---under measurement misalignment. In the binary-outcome, uniform-marginal setting, all three can attain the same peak extractable work, $k_{\mathrm B}T\ln 2$. Their operational value differs, however, in robustness away from perfect alignment. For the chosen classical benchmark the mutual information degrades as $\Theta(\delta\theta\ln(1/\delta\theta))$, whereas for the quantum cosine  law it degrades as $\Theta(\delta\theta^2\ln(1/\delta\theta))$. The stronger-than-quantum step law is perfectly flat except at a critical angle. Accordingly, the paper establishes a robustness hierarchy \emph{within this restricted comparison class}, rather than a no-go theorem for all classical correlations.}
\end{abstract}

\maketitle

\section{{Introduction}}

Classical thermodynamics defines an isolated system by boundaries
impermeable to matter and energy.
This picture is challenged by microscopic systems prepared with initial
correlations to their environment.
Even when sealed against energy exchange, such a system's local state
can be altered by operations on its correlated environmental
counterpart.
An agent possessing side information can consume these correlations to
perform work, a process governed by generalized thermodynamic laws that
incorporate information-theoretic
balances~\cite{Bera2017a,delrio2011thermodynamic,Sagawa2012}.

{The connection between information and thermodynamics has a long
history, originating with Maxwell's ``demon'' thought experiment and
formalized through the work of Szilard~\cite{Szilard-1929,maxwell-demon2}, who showed
that a single bit of information about a molecule's position in a
partitioned box can be converted into $k_{\mathrm{B}}T\ln 2$ of work.
Landauer~\cite{landauer:61} later established that this cycle is closed
by the erasure cost: resetting the demon's memory dissipates at least
$k_{\mathrm{B}}T\ln 2$ of heat, thereby saving the second
law~\cite{bennett-82,Maruyama2009}.}

{In the quantum regime, information-thermodynamic trade-offs are enriched
by genuinely non-classical correlations.
It has been shown that quantum
discord~\cite{Zurek2003,Oppenheim2002} and quantum
entanglement~\cite{delrio2011thermodynamic} can serve as thermodynamic
resources, and that the framework of generalized free
energies~\cite{Brandao2015,Brandao2015b,Horodecki2013} extends the
second law to the single-shot regime.
Sagawa and Ueda~\cite{SagawaUeda2008,sagawa2010} established
rigorous equalities connecting feedback control, measurement,
and work extraction in the quantum setting.
Comprehensive reviews of the field can be found in
Refs.~\cite{Goold2016,Vinjanampathy2016}.}

{This paper compares three representative correlation laws and shows that,
within a natural singlet-like comparison class, their peak energetic
value can coincide while their robustness against misalignment differs
sharply.
For the three representative singlet-like laws studied here, greater
nonlocality coincides with greater thermodynamic robustness under
misalignment.
This robustness hierarchy is quantified by an aggregate thermodynamic
parameter inspired by the measurement settings of the
Clauser--Horne--Shimony--Holt (CHSH)
scenario~\cite{chsh,bell,Brunner2014}, though this parameter is used as
an operational figure of merit rather than as a Bell inequality.}

{The remainder of this paper is organized as follows.
Section~\ref{sec:prelim} reviews the information-theoretic framework
connecting mutual information to extractable work and derives the mutual
information for binary-outcome correlations.
Section~\ref{sec:protocol} describes the operational
work-extraction protocol.
Section~\ref{sec:hierarchy} introduces the three representative
correlation laws.
Section~\ref{sec:thermo_value} analyzes their thermodynamic value and
robustness, including a Bell-inequality argument that constrains the
small-angle behavior of local singlet-like models.
Section~\ref{sec:CHSH_work} defines and evaluates an aggregate
thermodynamic robustness parameter across multiple measurement
settings.
Section~\ref{sec:extensions} discusses extensions and limitations.
Section~\ref{sec:discussion} positions the work relative to the
existing literature.
Section~\ref{sec:conclusion} concludes.}

\section{{Preliminaries}}
\label{sec:prelim}

\subsection{{From information to work}}
\label{sec:info_to_work}

{Consider a bipartite system consisting of a subsystem~$A$ and its
environment~$E$, jointly described by a state $\rho_{AE}$ with reduced
states $\rho_A = \mathrm{Tr}_E[\rho_{AE}]$ and
$\rho_E = \mathrm{Tr}_A[\rho_{AE}]$.
The quantum mutual information,
\begin{equation}
I(A:E) = S(\rho_A) + S(\rho_E) - S(\rho_{AE}),
\label{eq:QMI}
\end{equation}
where $S(\rho) = -\mathrm{Tr}[\rho \ln \rho]$ is the von~Neumann
entropy, quantifies the total correlations (both classical and quantum)
between $A$ and~$E$.}

{If $A$ is in contact with a thermal bath at temperature $T$, and the
agent performs a cyclic process on $A$ that begins with correlations
$I_i(A{:}E)$ and terminates in an uncorrelated product state
$\rho_A\otimes\rho_E$, the generalized second
law~\cite{Bera2017a} imposes the following bound on the extractable
work:
\begin{equation}
W_{\mathrm{ext}} \le -\Delta F_T(\rho_A)
  + k_{\mathrm{B}} T\, I_i(A{:}E),
\label{eq:gen_second_law}
\end{equation}
where $\Delta F_T(\rho_A)$ is the change in the non-equilibrium free
energy of subsystem~$A$.
This result can be understood intuitively: the mutual information
$I_i(A{:}E)$ represents a bonus resource, additional to the
system's own free energy, that an informed agent can exploit.
When the process is cyclic on~$A$, so that $\Delta F_T(\rho_A)=0$,
this simplifies to}
\begin{equation}
W_{\mathrm{ext}} \le k_{\mathrm{B}} T\, I_i(A{:}E).
\label{eq:work_from_I}
\end{equation}
{The physical content of Eq.~\eqref{eq:work_from_I} generalizes
Szilard's result~\cite{Szilard-1929}: one bit of mutual information
($I = \ln 2$~nats) can be reversibly converted into
$k_{\mathrm{B}}T\ln 2$ of work, precisely the energy liberated by an
isothermal expansion of a single-molecule gas when the piston's initial
position is known.
Following del~Rio~\emph{et al.}~\cite{delrio2011thermodynamic}, the
work cost of erasing a system is governed by the conditional entropy
$S(A|E) = S(\rho_{AE}) - S(\rho_E)$, which can be negative for
entangled states, permitting work \emph{gain} during erasure.
The work bound in Eq.~\eqref{eq:work_from_I} is consistent with this
picture, since for a maximally mixed marginal $\rho_A$ (entropy
$S(\rho_A) = \ln 2$), we have $I(A{:}E) = \ln 2 - S(A|E)$.}

{\emph{Scope of the bound.}
The quantum mutual information of Eq.~\eqref{eq:QMI} can exceed
$\ln 2$ for quantum subsystems (for instance,
$I(A{:}E) = 2\ln 2$ for a maximally entangled two-qubit state),
and in principle more general protocols could exploit such larger
correlations~\cite{delrio2011thermodynamic}.
In the present work we restrict attention to the
\emph{outcome-level} mutual information arising from binary
measurements with uniform marginals, for which
$I \le \ln 2$ always holds.
The bound $W \le k_{\mathrm{B}}T \ln 2$ is therefore specific to
this one-unbiased-bit operational setting, not universal across all
quantum thermodynamic protocols.}

\subsection{{Mutual information from correlations}}
\label{sec:MI_from_E}

{We now specialize to a scenario directly relevant to Bell-type
experiments.
Consider two parties sharing a no-signaling bipartite system with binary
outcomes $a, e \in \{+1, -1\}$ and uniform marginals:
$P(a) = P(e) = 1/2$.
In this case, the joint distribution is entirely determined by the
correlation function}
\begin{equation}
{E(\theta)
  = \langle a\, e \rangle_\theta
  = \sum_{a,e} a\, e\, P(a,e|\theta),}
\label{eq:corr_def}
\end{equation}
{where $\theta$ is the relative angle between the two measurement
settings.
Explicitly, the joint probabilities are
\begin{equation}
P(a = e|\theta)   = \frac{1+E(\theta)}{2},
\qquad
P(a \neq e|\theta) = \frac{1-E(\theta)}{2}.
\label{eq:joint_probs}
\end{equation}
The marginal entropy is $H(A) = \ln 2$.
The conditional entropy is the binary entropy of the error probability,
\begin{equation}
H(A|E)
  = h_2\!\left(\frac{1-E(\theta)}{2}\right)
  = h_2\!\left(\frac{1+E(\theta)}{2}\right),
\end{equation}
where $h_2(p) = -p\ln p - (1{-}p)\ln(1{-}p)$ is the binary Shannon
entropy in nats, and the last equality follows from the symmetry
$h_2(p)=h_2(1{-}p)$.
The mutual information is therefore}
\begin{equation}
I_\theta(A{:}E) = {H(A) - H(A|E) =} \ln 2
  - h_2\!\left(\frac{1+E(\theta)}{2}\right).
\label{eq:Itheta}
\end{equation}
{This result depends on the correlation $E(\theta)$ only through
its absolute value $|E(\theta)|$, via the binary entropy.
Perfect correlation or anti-correlation ($|E|=1$) gives $I = \ln 2$;
zero correlation ($E=0$) gives $I = 0$.
Equation~\eqref{eq:Itheta} is model-independent: it holds for any
bipartite device (or theory) whose outputs are binary with uniform
marginals, in which case the joint distribution is fully determined by
the correlator $E(\theta)=\langle AE\rangle_\theta$.}

\section{{Operational Protocol}}
\label{sec:protocol}

{To make the work-extraction scenario concrete, we describe an
operational protocol inspired by the Szilard
engine~\cite{Szilard-1929,SagawaUeda2008}.
We treat the measurement outcomes as classical registers; hence
$I_\theta(A{:}E)$ denotes the Shannon mutual information of the binary
variables $A,E\in\{\pm1\}$.}

\begin{enumerate}
\item {Alice and Bob share a correlated resource (classical shared
randomness, an entangled quantum state, or a no-signaling box).}

\item {A referee assigns measurement settings: Alice receives setting $x$
(direction $\alpha_x$) and Bob receives setting $y$ (direction $\beta_y$).
The settings (equivalently, the relative angle
$\theta=\alpha_x-\beta_y$) are assumed to be known to both parties
(e.g., publicly announced by the referee).}

\item {Alice performs her measurement and obtains $a\in\{+1,-1\}$.
Bob queries his device with setting $y$, obtaining $e\in\{+1,-1\}$,
which is stored in a physical two-state working medium
(e.g., Left/Right of a one-molecule Szilard box at temperature $T$).}

\item {Alice communicates her one-bit result $a$ to Bob via a classical
channel.}

\item {Using $a$ and the known correlator $E(\theta)=\langle AE\rangle_\theta$,
Bob performs a reversible relabeling (a conditional flip) of his
two-state working medium and defines an agreement variable
\[
z := \sigma(\theta)\, a e \in \{+1,-1\},
\]
where $\sigma(\theta)\in\{+1,-1\}$ is chosen to maximize
$P(z=+1|\theta)$
(equivalently, $\sigma(\theta)=\mathrm{sgn}\,E(\theta)$ whenever
$E(\theta)\neq 0$).
For uniform marginals,
\begin{equation}
\begin{split}
p_{\rm corr}(\theta) &= P(z=+1|\theta)  \\
&= \max\!\Big\{\frac{1+E(\theta)}{2},\frac{1-E(\theta)}{2}\Big\} \\
&= \frac{1+|E(\theta)|}{2}\,.
\end{split}
\label{eq:pcorr}
\end{equation}}

\item {Bob now performs a feedback-controlled quasi-static isothermal
protocol that brings his two-state working medium from the biased
distribution $P(z)$ to thermal equilibrium (uniform).
The $A$--$E$ correlation is the informational resource being consumed.
The maximum average extractable work per run is~\cite{SagawaUeda2008,parrondo2015}
\begin{equation}
W(\theta) = k_{\mathrm{B}}T\,I_\theta(A{:}E)
= k_{\mathrm{B}}T\!\left[\ln 2
- h_2\!\left(\frac{1+|E(\theta)|}{2}\right)\right]\!.
\label{eq:wtheta}
\end{equation}}
\end{enumerate}

{Since Alice communicates exactly one classical bit and the marginals are
uniform, the mutual information satisfies
$I_\theta(A{:}E) \le H(A) = \ln 2$, hence
$W(\theta)\le k_{\mathrm{B}}T\ln 2$ in this one-unbiased-bit
setting~\cite{landauer:61,bennett-82}.
This bound is saturated when $|E(\theta)|=1$, i.e., when Bob can
deterministically predict (equivalently, deterministically relabel) the
logical state of his two-state working medium.}

{\paragraph{Thermodynamic bookkeeping.}
Equation~\eqref{eq:wtheta} quantifies the \emph{gross} work that can be
extracted from Bob's two-state working medium given access to Alice's
one-bit side information.
The energetic cost of creating, transmitting, storing, and eventually
resetting the communication record is not included in $W(\theta)$.
Accordingly, our analysis compares the conditional work value of
different correlation laws under a common one-bit feedforward resource;
it does not claim a positive \emph{net} work gain for a fully closed
engine cycle once communication and memory-reset costs are included.}

\section{A Hierarchy of Correlation Laws}
\label{sec:hierarchy}

{Our aim is not to classify all local-hidden-variable correlations.
Without further restrictions, trivial local strategies with unbiased
marginals can produce angle-independent perfect anti-correlation, e.g.
$E(\theta)\equiv -1$, and therefore a flat work profile.
Such strategies do not model a singlet-like angular response and are not
part of the present comparison.
Throughout, we restrict attention to singlet-like,
angle-dependent, rotationally covariant correlation laws with unbiased
marginals satisfying
\begin{equation}
E(0)=-1,\qquad E(\pi)=+1,\qquad E(\pi-\theta)=-E(\theta),
\label{eq:singlet_symmetry}
\end{equation}
together with monotonic increase on $[0,\pi]$.
Within this restricted class, the linear local model serves as a
classical benchmark.}

We analyze three representative bipartite correlation laws $E(\theta)$
for $\{\pm1\}$-outcome measurements as a function of the relative
angle $\theta \in [0, \pi]$ between settings.
Their nonlocality is gauged by the {CHSH}
parameter{~\cite{chsh,bell}}
\begin{equation}
\mathcal{S}_{\mathrm{CHSH}}
  {= |E(\theta_1,\theta_2) + E(\theta_1,\theta_2')
    + E(\theta_1',\theta_2) - E(\theta_1',\theta_2')|,}
\end{equation}
for which local realism imposes a bound of~2, quantum mechanics a
bound of $2\sqrt{2}$ (the Tsirelson
bound~\cite{cirelson:80}), and the no-signaling principle an algebraic
limit of~4.
The three models are (see Fig.~\ref{fig:correlations}):
\begin{enumerate}
  \item \textbf{Classical {benchmark} (linear {/ ``triangle'' model}):}
  \[
    E_c(\theta) = -1 + {\frac{2\theta}{\pi}\qquad (0\le \theta \le \pi).}
  \]
  {With the extension to other angles understood by the usual symmetries,
  this is a standard rotationally covariant local-hidden-variable
  correlation satisfying the singlet-like endpoint conditions
  $E(0)=-1$ and $E(\pi)=+1$ and providing a simple classical benchmark
  for angle-induced degradation.
  As shown in Sec.~\ref{sec:robustness}, Bell's original inequality
  implies that within the restricted local class the small-angle
  behavior is necessarily at best linear-cusp-like, and the triangle
  model saturates the resulting bounds.
  For the conventional CHSH choice of settings used in this work, it
  attains the local bound $\mathcal{S}_{\mathrm{CHSH}}^{(c)}=2$.}

  \item \textbf{Quantum (cosine):}
      \[
      E_q(\theta) = -\cos\theta.
      \]
      This describes measurements on a two-qubit singlet state and
      reaches the Tsirelson bound,
      $\mathcal{S}_{\mathrm{CHSH}}^{(q)} = 2\sqrt{2}$.

  \item \textbf{Stronger-than-quantum (step):}
      {\[
      E_s(\theta)=
      \begin{cases}
      -1, & 0\le \theta<\pi/2,\\
      0,  & \theta=\pi/2,\\
      +1, & \pi/2<\theta\le \pi.
      \end{cases}
      \]}
      This hypothetical box correlation saturates the algebraic
      limit,
      $\mathcal{S}_{\mathrm{CHSH}}^{(s)} = 4$~\cite{pop-rohr,%
      svozil-krenn}.
      {The standard Popescu--Rohrlich box is defined for finitely many
      settings; the extension to a continuous function $E(\theta)$ is an
      additional idealization consistent with no-signaling but not
      uniquely fixed by the PR-box definition.}
\end{enumerate}

\begin{figure}[ht!]
    \centering
\resizebox{0.4\textwidth}{!}{%
\begin{tikzpicture}
   \begin{axis}[
        xlabel={Relative Angle $\theta$},
        ylabel={Correlation $E(\theta)$},
        xmin=0, xmax=180, ymin=-1.1, ymax=1.1,
        axis lines=left,
        xtick={0, 45, 90, 135, 180},
        xticklabels={$0$, $\pi/4$, $\pi/2$, $3\pi/4$, $\pi$},
        ytick={-1, 0, 1},
        legend style={at={(0.01,0.93)},
          anchor=north west, font=\small},
        legend cell align={left},
        grid=major, width=10cm, height=7cm,
    ]
    \addplot[domain=0:180, samples=200,
      color=red, ultra thick]
      { -cos(x) };
    \addlegendentry{Quantum ($E_q$)}
    \addplot[domain=0:180, samples=2,
      color=green!70!black, dashed, ultra thick]
      { -1 + 2*x/180 };
    \addlegendentry{Classical {benchmark} ($E_c$)}
    \addplot[domain=0:90, samples=2,
      color=blue, dotted, ultra thick]
      { -1 };
    \addlegendentry{Stronger-than-quantum ($E_s$)}
    \addplot[domain=90:180, samples=2,
      color=blue, dotted, ultra thick, forget plot]
      { 1 };
    \end{axis}
\end{tikzpicture}
}
    \caption{{Comparison of the three representative correlation
    functions $E(\theta)$.
    The quantum (solid red) and classical benchmark (dashed green)
    only achieve perfect anti-correlation at $\theta=0$.
    The stronger-than-quantum correlation (dotted blue) is perfect for
    all angles except $\theta = \pi/2$.}}
    \label{fig:correlations}
\end{figure}

\section{Thermodynamic Value and Robustness}
\label{sec:thermo_value}

\subsection{{Peak work and mutual information}}

The thermodynamic value of these correlations is {governed} by the mutual
information $I_\theta(A{:}E)$ {derived in Eq.~\eqref{eq:Itheta}}.
As shown in Fig.~\ref{fig:mutual_information}, perfect correlation or
anti-correlation ($|E(\theta)|=1$) yields the maximal mutual
information $I=\ln 2$.
Consequently, according to Eq.~\eqref{eq:work_from_I}, the maximum
work extractable per bipartite system {in this binary-outcome,
uniform-marginal setting} is bounded by
$k_{\mathrm{B}}T\ln 2$.
This peak value is achievable in all three {representative} models, for
instance, by aligning the measurement bases ($\theta=0$).

\begin{figure}[ht!]
\pgfmathdeclarefunction{h2}{1}{%
  \pgfmathparse{%
    -((#1)==0 ? 0 : (#1)*ln(#1))
    - ((1-#1)==0 ? 0 : (1-#1)*ln(1-#1))}%
}
\resizebox{0.4\textwidth}{!}{%
    \begin{tikzpicture}
        \begin{axis}[
            xlabel={Relative Angle $\theta$},
            ylabel={Mutual Information $I(\theta)$ (nats)},
            xmin=0, xmax=180, ymin=-0.05, ymax=0.75,
            axis lines=left,
            xtick={0, 45, 90, 135, 180},
            xticklabels={$0$, $\pi/4$, $\pi/2$,
              $3\pi/4$, $\pi$},
            ytick={0, 0.6931},
            yticklabels={0, $\ln 2$},
            legend style={at={(0.5, 0.75)},
              anchor=center, font=\small},
            legend cell align={left},
            grid=major, width=10cm, height=7cm,
        ]
        \addplot[domain=0:180, samples=201,
          color=red, ultra thick]
          { ln(2) - h2( (sin(x/2))^2 ) };
        \addlegendentry{Quantum ($I_q$)}
        \addplot[domain=0:180, samples=101,
          color=green!70!black, ultra thick, dashed]
          { ln(2) - h2(x/180) };
        \addlegendentry{Classical {benchmark} ($I_c$)}
        \addplot[domain=0:180, samples=2,
          color=blue, dotted, ultra thick]
          { ln(2) };
        \addlegendentry{Stronger-than-quantum ($I_s$)}
        \addplot[only marks, mark=*, mark size=3pt,
          color=blue, fill=white]
          coordinates {(90, {ln(2)})};
        \addplot[only marks, mark=*, mark size=2pt,
          color=blue, fill=blue]
          coordinates {(90, 0)};
        \end{axis}
    \end{tikzpicture}
}
    \caption{{Mutual information versus measurement angle $\theta$.
    All three representative models achieve the maximum value $\ln 2$,
    but their robustness to misalignment differs.
    The classical benchmark (dashed green) degrades as
    $\Theta(\delta\theta\ln(1/\delta\theta))$ from the peak, while
    the quantum cosine law (solid red) degrades as
    $\Theta(\delta\theta^{2}\ln(1/\delta\theta))$, offering superior
    resilience within the restricted comparison class.}}
    \label{fig:mutual_information}
\end{figure}

\subsection{Robustness analysis}
\label{sec:robustness}

This thermodynamic equivalence in peak yield belies a critical
operational difference: robustness.
The practical advantage of non-classical correlations becomes apparent
when considering misalignment from the optimal setting ($\theta=0$).

\paragraph{Classical {benchmark (linear)} robustness.}
{For the classical benchmark,
$E_c(\delta\theta) = -1 + 2\delta\theta / \pi$ near $\theta = 0$.
Setting $p = (1 + E_c)/2 = \delta\theta / \pi$, the binary entropy
for small $p$ behaves as
$h_2(p) \approx -p \ln p + p + \mathcal{O}(p^2)$.
Therefore,
\begin{equation}
I_c(\delta\theta)
  = \ln 2 - h_2\!\left(\frac{\delta\theta}{\pi}\right)
  \approx \ln 2
    - \frac{\delta\theta}{\pi}
      \ln\!\frac{\pi}{\delta\theta}
    - \frac{\delta\theta}{\pi}
    + \cdots .
\label{eq:Ic_expansion}
\end{equation}
The leading correction is
$\Theta(\delta\theta\ln(1/\delta\theta))$, worse than linear:
the mutual information degrades sharply for any nonzero misalignment.
This makes the chosen classical benchmark brittle with respect to
experimental noise.}

{\paragraph{Why the local benchmark is generic within the restricted class.}
The use of a linear classical benchmark is not arbitrary.
For local hidden-variable models with perfect anti-correlation,
Bell's original inequality~\cite{bell} implies
\begin{equation}
|E(\mathbf a,\mathbf b)-E(\mathbf a,\mathbf c)|
\le 1+E(\mathbf b,\mathbf c).
\label{eq:bell_original}
\end{equation}
For rotationally covariant coplanar correlations
$E(\mathbf a,\mathbf b)=E(|\alpha-\beta|)$, choose settings with
$|\alpha-\beta|=2\theta$ and
$|\alpha-\gamma|=|\beta-\gamma|=\theta$.
Then
\begin{equation}
|E(2\theta)-E(\theta)| \le 1+E(\theta).
\label{eq:bell_isotropic}
\end{equation}
Since the restricted comparison class assumes monotonic increase on
$[0,\pi]$, the left-hand side equals $E(2\theta)-E(\theta)$, and
therefore
\begin{equation}
1+E(2\theta)\le 2\,[1+E(\theta)].
\label{eq:delta_bound}
\end{equation}
Now suppose that near $\theta=0$ one has
$1+E(\theta)\sim c\,\theta^n$ with $c>0$.
Equation~\eqref{eq:delta_bound} then implies
$2^n\le 2$, hence
\begin{equation}
n\le 1.
\label{eq:n_le_1}
\end{equation}
Thus no local rotationally covariant singlet-like model in the present
comparison class can approach $-1$ quadratically at the origin; its
small-angle behavior is necessarily at best linear-cusp-like.
The triangle model saturates this constraint.}

{The same inequality constrains the CHSH angle used later.
Writing $\delta(\theta):=1+E(\theta)$, Eq.~\eqref{eq:delta_bound} gives
$\delta(\pi/2)\le 2\delta(\pi/4)$ and
$\delta(\pi)\le 2\delta(\pi/2)$, hence
\[
2=\delta(\pi)\le 4\delta(\pi/4).
\]
Therefore $\delta(\pi/4)\ge 1/2$.
Using the singlet symmetry $E(\pi-\theta)=-E(\theta)$, we have
$E(\pi/2)=0$ and thus $E(\pi/4)\le 0$, so
\begin{equation}
|E(\pi/4)| = -E(\pi/4) \le \frac12.
\label{eq:local_pi_over_4_bound}
\end{equation}
Within the restricted local comparison class,
\begin{equation}
\mathcal S_W^{\rm(local)}
\le 4k_{\mathrm B}T\!\left[\ln2-h_2\!\left(\tfrac14\right)\right],
\label{eq:SW_local_bound}
\end{equation}
and the triangle model saturates this bound.}

\paragraph{Quantum (cosine) robustness.}
{For the quantum model,
$E_q(\delta\theta) = -\cos\delta\theta
  \approx -1 + \delta\theta^2/2$.
Setting
$p = (1 + E_q)/2 \approx \delta\theta^2/4$, we have
$h_2(p) \approx -p\ln p + p
       \approx \frac{\delta\theta^2}{4}
              \ln\!\frac{4}{\delta\theta^2}
             + \frac{\delta\theta^2}{4}$
for small $\delta\theta$.
More precisely, expanding Eq.~\eqref{eq:Itheta} with
$E_q = -\cos\theta$ we can write
\begin{equation}
I_q(\delta\theta)
  = \ln 2 - h_2\!\left(\sin^2\!\frac{\delta\theta}{2}\right)
  \approx \ln 2
    - \Theta(\delta\theta^2 \ln(1/\delta\theta)).
\label{eq:Iq_expansion}
\end{equation}
The leading correction is quadratic up to logarithmic factors.
The quantum correlation $E_q(\theta) = -\cos\theta$ has a stationary
point at $\theta = 0$,
which is the fundamental reason for this improved robustness.}
Within the {present comparison class}, the quantum resource is
significantly more {resilient} against small imperfections in
establishing a shared reference frame.

\paragraph{Stronger-than-quantum (step) robustness.}
The stronger-than-quantum model {provides the ultimate in robustness:
for any $\delta\theta \neq \pi/2$, we have
$|E_s(\delta\theta)| = 1$ and therefore
$I_s(\delta\theta) = \ln 2$.
Coordination} of measurement settings is almost entirely unnecessary{;
the resource is maximal for all but a single critical angle.}

{\paragraph{General criterion.}
The scaling hierarchy above is not an accident of the three specific
models: once a correlation law is specified, the robustness of the
mutual information near a peak follows from the leading nonzero term in
the expansion of $1+E(\theta)$ around the extremum.
If
\[
E(\theta) = -1 + c\,|\theta|^n + o(|\theta|^n)
\qquad (c>0,\; n>0),
\]
then the binary entropy implies
\begin{equation}
\ln 2 - I(\theta)
= \Theta(|\theta|^n \ln(1/|\theta|)).
\label{eq:general_scaling}
\end{equation}
In particular, a linear cusp produces
$\Theta(\delta\theta\ln(1/\delta\theta))$ degradation, whereas a
stationary quadratic extremum produces
$\Theta(\delta\theta^2\ln(1/\delta\theta))$ degradation.
Bell's original inequality shows that, within the restricted local
singlet-like class considered here, one must have $n\le 1$.
The quantum cosine law instead has $n=2$.
Thus the thermodynamic advantage of the quantum example in this paper is
not tied to the singlet state alone, but to the presence of a smooth
stationary extremum.
Table~\ref{tab:summary} summarizes the key properties of the three
representative models.}

\begin{table*}[t]
\caption{\label{tab:summary}
{Summary of the three representative correlation laws.
$\mathcal{S}_{\mathrm{CHSH}}$ is the CHSH value;
$W_{\max}$ is the peak extractable work
(in units of $k_{\mathrm{B}}T\ln 2$);
the ``Degradation of~$I$'' column gives the leading magnitude of the
correction to
$I(\delta\theta) = \ln 2 - \cdots$ for small misalignment
$\delta\theta$.}}
\begin{ruledtabular}
\begin{tabular}{lccc}
{Class} & {$\mathcal{S}_{\mathrm{CHSH}}$}
      & {$W_{\max}/k_{\mathrm{B}}T\ln 2$}
      & {Degradation of $I$ near $\theta{=}0$} \\
\midrule
{Classical benchmark} & {$2$}
  & {$1$} & {$\Theta(\delta\theta\ln(1/\delta\theta))$} \\
{Quantum} & {$2\sqrt{2}$}
  & {$1$} & {$\Theta(\delta\theta^{2}\ln(1/\delta\theta))$} \\
{Stronger-than-quantum} & {$4$}
  & {$1$} & {$0$ \;(perfectly robust)} \\
\end{tabular}
\end{ruledtabular}
\end{table*}

\section{{Multi-Setting Thermodynamic Robustness}}
\label{sec:CHSH_work}

{The operational hierarchy established above can be quantified by
an aggregate robustness parameter that evaluates the total
thermodynamic value across multiple measurement settings, as would
arise in a protocol where the measurement angle varies across runs.}

{We adopt the standard CHSH measurement configuration---Alice's
settings $\alpha_1 = 0$, $\alpha_2 = \pi/2$ and Bob's settings
$\beta_1 = \pi/4$, $\beta_2 = -\pi/4$---which produces four
setting pairs with relative angles
$\theta_{11} = \pi/4$, $\theta_{12} = \pi/4$,
$\theta_{21} = \pi/4$, $\theta_{22} = 3\pi/4$.
For isotropic correlations satisfying $|E(\theta)| = |E(\pi-\theta)|$,
one has $I(\pi/4) = I(3\pi/4)$, so the four terms all take the same
value.}

{We define the aggregate thermodynamic robustness parameter
\begin{equation}
\mathcal{S}_W
  = k_{\mathrm{B}}T\sum_{k=1}^{4}
    I(\theta_k)
  = k_{\mathrm{B}}T\bigl[3\,I(\pi/4) + I(3\pi/4)\bigr].
\label{eq:SW_def}
\end{equation}}

{\emph{Important remark.}
Although $\mathcal{S}_W$ is evaluated at the CHSH measurement angles,
it is \emph{not} a Bell inequality: no local bound is derived or
claimed from sign-sensitive CHSH algebra.
The standard CHSH functional depends on the \emph{signs} of the
correlators through the combination
$E + E + E - E$, which is what enables the separation between local
and nonlocal theories.
The mutual information $I(\theta)$ depends only on $|E(\theta)|$
and is therefore insensitive to the sign structure.
We use $\mathcal{S}_W$ purely as an operational figure of merit
quantifying how much total work can be extracted when measurement
settings span the CHSH configuration.}

{For the restricted local comparison class,
Eq.~\eqref{eq:local_pi_over_4_bound} already implies the bound
\eqref{eq:SW_local_bound}.
The triangle model saturates that local bound.
We now evaluate $\mathcal{S}_W$ explicitly for each of the three
representative models.}

{\paragraph{Classical benchmark.}
Using Eq.~\eqref{eq:Itheta} with $E_c(\pi/4) = -1/2$ and
$E_c(3\pi/4) = 1/2$, so that $|E| = 1/2$ in both cases,
we obtain $I(\pi/4) = I(3\pi/4) = \ln 2 - h_2(1/4)$.
Computing $h_2(1/4)$ explicitly:
\begin{align}
h_2\left(\tfrac{1}{4}\right) &= -\tfrac{1}{4}\ln\tfrac{1}{4}
             -\tfrac{3}{4}\ln\tfrac{3}{4} \notag\\
          &= 2\ln 2 - \tfrac{3}{4}\ln 3
           \approx 0.5623 \;\text{nats}.
\label{eq:h2_quarter}
\end{align}
Therefore
$I(\pi/4) = \ln 2 - (2\ln 2 - \tfrac{3}{4}\ln 3)
           = \tfrac{3}{4}\ln 3 - \ln 2
           \approx 0.1308$~nats, and
\begin{equation}
\mathcal{S}_W^{(c)}
  = 4 \times 0.1308\, k_{\mathrm{B}}T
  \approx 0.523\, k_{\mathrm{B}}T.
\label{eq:SW_classical}
\end{equation}}

{\paragraph{Quantum.}
With $E_q(\pi/4) = -\cos(\pi/4) = -1/\sqrt{2}$ and
$E_q(3\pi/4) = -\cos(3\pi/4) = 1/\sqrt{2}$, we similarly have
$|E| = 1/\sqrt{2}$ and
$I(\pi/4) = I(3\pi/4)
  = \ln 2 - h_2\!\bigl(\tfrac{1+1/\sqrt{2}}{2}\bigr)$,
where $p = (1-1/\sqrt{2})/2 = \sin^2(\pi/8) \approx 0.1464$.
Computing $h_2(0.1464)$ directly:
\begin{align}
h_2(0.1464) &= -0.1464\ln(0.1464) - 0.8536\ln(0.8536)
\notag\\
  &\approx 0.2813 + 0.1352 \notag\\
  &\approx 0.4165~\text{nats}.
\label{eq:h2_quantum}
\end{align}
Therefore $I(\pi/4) \approx 0.6931 - 0.4165
  = 0.2766$~nats, and
\begin{equation}
\mathcal{S}_W^{(q)}
  = 4 \times 0.2766\, k_{\mathrm{B}}T
  \approx 1.107\, k_{\mathrm{B}}T.
\label{eq:SW_quantum}
\end{equation}}

{\paragraph{Stronger-than-quantum.}
With $|E_s(\pi/4)| = |E_s(3\pi/4)| = 1$, we have
$I(\pi/4) = I(3\pi/4) = \ln 2 \approx 0.6931$~nats.
Therefore,
\begin{equation}
\mathcal{S}_W^{(s)}
  = 4\,k_{\mathrm{B}}T\ln 2
  \approx 2.773\, k_{\mathrm{B}}T.
\label{eq:SW_superquantum}
\end{equation}}

{These values exhibit the strict hierarchy
\begin{equation}
\mathcal{S}_W^{(c)} < \mathcal{S}_W^{(q)}
  < \mathcal{S}_W^{(s)}.
\end{equation}
For the three representative laws studied here, stronger nonlocality
coincides with a larger aggregate thermodynamic value across diverse
measurement settings.}

\section{{Extensions and Limitations}}
\label{sec:extensions}

\subsection{{Relaxing the uniform-marginal assumption}}
\label{sec:non_uniform}

{The derivation of the mutual information in Eq.~\eqref{eq:Itheta}
relied on two structural assumptions: (i) binary outcomes and
(ii) uniform marginals ($P(a) = P(e) = 1/2$).
Assumption~(ii) can be relaxed straightforwardly at the level of
definitions.
For a general binary no-signaling distribution with marginals
$P(a{=}+1) = q_A$ and $P(e{=}+1) = q_E$, the mutual information is
\begin{equation}
I(A{:}E) = H(A) + H(E) - H(A,E),
\label{eq:I_general_binary}
\end{equation}
where all entropies are Shannon entropies of the respective
distributions.
With non-uniform marginals, however, the joint distribution is no longer
determined by $E(\theta)$ alone: one additionally needs the marginals.
The work bound~\eqref{eq:work_from_I} remains valid, but the simple
formula~\eqref{eq:Itheta} is replaced by an expression involving the
full binary table.}

{For binary outcomes one still has
$I(A{:}E) \le \min\{H(A), H(E)\} \le \ln 2$, so the peak
extractable work remains at most $k_{\mathrm{B}}T\ln 2$.
But the present robustness comparison is specific to the binary,
uniform-marginal setting, where $I(\theta)$ is determined solely by
$E(\theta)$.
Extensions beyond that setting require additional assumptions about the
full joint distribution and are not pursued here.}

\subsection{{Beyond binary outcomes}}
\label{sec:beyond_binary}

{For $d$-outcome measurements, the joint distribution is no longer
captured by a single scalar correlation function, and the conditional
entropy involves the full $d\times d$ probability table.
The extractable work is still bounded by
$k_{\mathrm{B}}T\, I(A{:}E)$, but the direct link to a single angular
correlator is lost.
Accordingly, the smoothness argument developed here does not transfer
automatically to $d>2$.
Extending the robustness analysis beyond the binary case is a natural
direction for future work.}

\subsection{{Single-shot corrections}}
\label{sec:single_shot}

{The work bound~\eqref{eq:work_from_I} is an asymptotic
many-copy result.
In the single-shot regime, the relevant quantities are suitable smooth
min- and max-entropies and the corresponding generalized free energies;
see, for example,
Refs.~\cite{Brandao2015b,Horodecki2013}.
A single-shot treatment of the present protocol would therefore replace
the Shannon/von~Neumann quantities used here by appropriate one-shot
entropic measures.
Whether the same robustness hierarchy persists in that regime is a
natural conjecture, but it is not established by the present analysis.}

\subsection{{Connection to device-independent protocols}}
\label{sec:DI}

{An analogous robustness advantage of quantum (and hypothetical
beyond-quantum) correlations appears in device-independent quantum key
distribution~\cite{acin2007} and certified randomness
expansion~\cite{acin2012}, where Bell-inequality violations can certify
security even when the devices are imperfect.
The thermodynamic robustness identified here is qualitatively similar:
within the restricted comparison class, smoother correlations degrade
more gracefully under misalignment.}

\section{{Discussion and Relation to Prior Work}}
\label{sec:discussion}

{Our results complement and extend several strands of the literature on
correlations as thermodynamic resources.}

{The foundational connection between correlations and work was
established by Oppenheim~\emph{et al.}~\cite{Oppenheim2002}, who
introduced the work deficit---the gap between globally and locally
extractable work---as a measure of quantum correlations.
While the work deficit quantifies how much correlation \emph{is}
present, the present analysis addresses a different question: how
robust a given correlation is as a fuel.
The two perspectives are complementary.}

{Zurek~\cite{Zurek2003} connected quantum discord to the work
extractable by Maxwell's demon, showing that the quantumness of
correlations has thermodynamic consequences.
The present work extends this perspective beyond quantum mechanics to
include an idealized no-signaling comparison point, and it identifies
robustness---rather than the absolute amount of extractable work---as
the distinguishing feature among the three representative laws
considered here.}

{Perarnau-Llobet~\emph{et al.}~\cite{PerarnauLlobet2015} provided a
comprehensive framework for extractable work from general correlated
states, proving tight bounds for local and global extraction protocols.
The present work specializes to a Bell-experiment-inspired scenario and
reveals a hierarchy in robustness that is not visible in statements
based solely on the total amount of correlation.}

{Funo \emph{et al.}~\cite{funo2013} showed that entanglement can
provide a thermodynamic advantage in work extraction under global
operations, while Hovhannisyan \emph{et al.}~\cite{hovhannisyan2013}
demonstrated that entanglement generation is not \emph{necessary} for
optimal work extraction by local operations.
The finding that $W_{\max} = k_{\mathrm{B}}T\ln 2$ for all three
representative laws is consistent with the latter observation:
in the present one-bit setting, the peak gross work is limited by the
communicated information, not by nonlocality itself.
The advantage of nonlocality here lies instead in the robustness with
which that peak can be approached.}

{Gallego \emph{et al.}~\cite{gallego2010} studied operational
advantages of nonlocal correlations in information-theoretic tasks.
The present work provides a thermodynamic analog: within the restricted
comparison class, the operational advantage is not a higher peak yield
but a greater tolerance to misalignment.}

{The resource-theoretic approach to
thermodynamics~\cite{Brandao2015,Brandao2015b,Horodecki2013}
introduces families of generalized free energies and establishes
second laws beyond the standard one.
The bound in Eq.~\eqref{eq:work_from_I} is the asymptotic
many-copy version of these constraints.
Single-shot corrections would alter the quantitative values of
$\mathcal{S}_W$, but their effect on the present robustness hierarchy
has not been analyzed here.}

{It is important to clarify the scope of these results.
The bound in Eq.~\eqref{eq:work_from_I} is standard within information
thermodynamics: for systems coupled to a heat bath at temperature $T$,
under the usual assumptions of the generalized second law, the maximum
average work extractable in a cyclic protocol is bounded by
$k_{\mathrm{B}}T$ times the appropriate mutual information.
In the operational setting studied here this is applied to the
classical outcome registers generated by the parties' measurements.
The expression for $I_\theta$ in Eq.~\eqref{eq:Itheta} is similarly
model-independent: it follows solely from binary outcomes, uniform
marginals, and the correlator $E(\theta)=\langle AE\rangle_\theta$.}

{What is model-dependent is the specific form of $E(\theta)$.
The linear function is a standard rotationally covariant local model and,
within the restricted singlet-like local class adopted here, Bell's
original inequality forces the small-angle behavior to be at best
linear-cusp-like and the CHSH-angle value to satisfy
$|E(\pi/4)|\le 1/2$; the triangle model saturates those bounds.
The cosine law arises from the singlet state, and the step law is an
idealized continuous no-signaling form reaching
$\mathcal{S}_{\mathrm{CHSH}}=4$~\cite{svozil-krenn}.
The robustness comparison should therefore be read as a statement about
this restricted comparison class, not as a classification of all
correlations permitted by a given theory.}

\section{{Conclusion}}
\label{sec:conclusion}

Correlations between a system and its environment are a thermodynamic
fuel whose value {in the present binary-outcome,
uniform-marginal setting} is limited by their mutual information.
A single perfectly correlated bit can yield at most
$k_{\mathrm{B}}T\ln 2$ of {gross} work, as exemplified by a Szilard engine
powered by shared information{~\cite{Szilard-1929}}.
Engines consuming correlations do not violate the second law; they
demonstrate that information must be treated as a physical
resource{~\cite{landauer:61,bennett-82}}.

{Within the restricted comparison class of representative singlet-like
correlation laws studied here, the hierarchy among classical, quantum,
and stronger-than-quantum correlations lies not in the peak value but in
the robustness with which it can be accessed.
The local singlet-like class is constrained by Bell's original
inequality to be at best linear-cusp-like near perfect alignment,
whereas the quantum cosine law has a smooth stationary extremum and is
therefore more resilient to small misalignment.
The idealized no-signaling step law is flatter still.
In this sense, quantum correlations are more robust than the classical
benchmark: not because they raise the peak gross yield in the present
one-bit setting, but because they preserve that yield more effectively
under misalignment. The idealized no-signaling step law is more robust
still.}

{The robustness hierarchy identified here should therefore be read as a
statement about a restricted comparison class of singlet-like,
rotationally covariant correlations, not as a theorem covering all
possible classical shared-randomness strategies.
Within that class, the quantum cosine law is markedly more tolerant to
misalignment than the classical linear benchmark, while the idealized
no-signaling step law is flatter still.}

\begin{acknowledgments}
This research was funded in whole or in part by the Austrian Science Fund (FWF) Grant DOI: 10.55776/PIN5424624. The author acknowledges TU Wien Bibliothek for financial support through its Open Access Funding Programme.
\end{acknowledgments}

\bibliography{svozil}

\end{document}